%% file: biokdd26.tex
\documentclass[sigconf]{acmart}
\AtBeginDocument{%
  }

\setcopyright{acmlicensed}
\copyrightyear{2026}
\acmYear{2026}
\acmDOI{XXXXXXX.XXXXXXX}
\acmConference[BioKDD'26]{25th International Workshop on Data Mining in Bioinformatics}{August 09--13, 2026}{Jeju, Korea}
\acmISBN{978-1-4503-XXXX-X/2018/06}

\usepackage[draft=false]{zebra-goodies}
\usepackage{booktabs}
\usepackage{dashrule}
\usepackage{array}
\usepackage{pifont}
\usepackage{multirow}
\usepackage{subcaption}
\usepackage{tikz}
\usepackage{xspace}
\usepackage{enumitem}
\usepackage{pgfplots}
\usepgfplotslibrary{colorbrewer}
\usepackage[most,breakable]{tcolorbox}
\tcbset{
  promptbox/.style={
    colback=blue!2!white,
    colframe=blue!60!black,
    boxrule=0.4pt, arc=2pt,
    left=8pt, right=8pt, top=6pt, bottom=6pt,
    fonttitle=\bfseries, breakable, enhanced
  }
}

\definecolor{Set2-A}{RGB}{102, 194, 165} 
\definecolor{Set2-B}{RGB}{252, 141, 98}  
\definecolor{Set2-C}{RGB}{141, 160, 203}
\definecolor{Set2-D}{RGB}{231, 138, 195}
\definecolor{Set2-E}{RGB}{166, 216, 84}

\newcommand{\mname}[0]{\textsc{LlmBdc}\xspace}

\begin{document}

\title[\mname: LLM for Gene Ontology Clustering]{\mname: \underline{L}anguage \underline{M}odel for \underline{B}iological \underline{D}omains Oriented \underline{C}lustering of Gene Ontology}

\author{Ximing Ran}
\affiliation{%
  \institution{Emory University}
  \city{Atlanta}
  \state{GA}
  \country{USA}}
\email{ximing.ran@emory.edu}

\author{Jie Xu}
\affiliation{%
  \institution{Emory University}
  \city{Atlanta}
  \state{GA}
  \country{USA}}
\email{jie.xu@emory.edu}

\author{Peng Jin}
\affiliation{%
  \institution{Emory University}
  \city{Atlanta}
  \state{GA}
  \country{USA}}
\email{peng.jin@emory.edu}

\author{Zhaohui Qin}
\affiliation{%
  \institution{Emory University}
  \city{Atlanta}
  \state{GA}
  \country{USA}}
\email{zhaohui.qin@emory.edu}

\author{Zhexing Wen}
\authornote{Corresponding authors: Zhexing Wen (zhexing.wen@emory.edu), Jiaying Lu (jiaying.lu@emory.edu)}
\affiliation{%
  \institution{Emory University}
  \city{Atlanta}
  \state{GA}
  \country{USA}}
\email{zhexing.wen@emory.edu}

\author{Jiaying Lu}
\authornotemark[1]
\affiliation{%
  \institution{Emory University}
  \city{Atlanta}
  \state{GA}
  \country{USA}}
\email{jiaying.lu@emory.edu}

\renewcommand{\shortauthors}{Ran et al.}

\begin{abstract}
Gene Ontology (GO) enrichment analysis is a foundational tool for translating large-scale genomic data into biological insights, but typically yields hundreds of redundant terms that obscure overarching themes.
Existing summarization tools rely on fixed similarity metrics (\texttt{REVIGO}, \texttt{GOSemSim}, \texttt{clusterProfiler::simplify()}), gene-overlap measures (\texttt{Metascape}), or static hierarchy mappings (\texttt{GO-slim}), and therefore cannot incorporate biological context. Manual curation provides context-aware grouping but is subjective and labor-intensive.
A scalable, context-aware framework is needed to cluster GO terms into interpretable higher-order biological domains.
Here we present \mname (Large Language Model for Biological Domains Oriented Clustering of Gene Ontology), a training-free framework that leverages zero-shot semantic reasoning of LLMs with confidence scoring to cluster GO terms into BioDomains using only ontology information at inference time.
Benchmarked across Alzheimer's disease (AD) and Fragile X syndrome (FXS) against six baseline methods including SapBERT, \mname achieved substantially higher precision, recall, and clustering performance. Against ground-truth annotations, \mname improved ARI from 9.7\% to 73.3\% (AD) and from 15.7\% to 66.6\% (FXS) over REVIGO, with corresponding NMI gains from 59.9\% to 73.4\% (AD) and 66.0\% to 79.5\%(FXS). A Cauchy combination test further confirmed that aggregated BioDomains retained statistically significant functional signals.
\mname provides a scalable, reproducible, and interpretable route to context-aware, system-level interpretation of GO enrichment results while preserving biological specificity.

\end{abstract}




\maketitle

\input{sections/intro}
\input{sections/related_work}
\input{sections/methods}

\input{sections/results}
\input{sections/discussion}
\input{sections/conclusion}


\bibliographystyle{ACM-Reference-Format}
\bibliography{mybib}


\end{document}

%% file: sections/intro.tex
\section{Introduction}
With the rise of high-throughput technologies, genomic data have become an increasingly important component of modern biological research, providing rich information on multi-layered gene regulation across a wide range of conditions, including disease states, developmental stages, and environmental perturbations~\cite{chen2016divan,wen2014synaptic}. A key step in interpreting these data is to understand the biological functions associated with molecular features of interest, as identified through various analytical approaches. Gene Ontology (GO) analysis~\cite{thomas2022panther} is one of the widely used approaches for this purpose. It identifies significantly overrepresented biological functions within a set of genes, providing a structured vocabulary to characterize biological processes, molecular functions, and cellular components -- collectively referred to as GO terms~\cite{ashburner2000GO}. These GO terms offer functional context for genes or gene-associated signals, helping researchers elucidate the mechanisms underlying the studied condition or phenotype.

To perform GO analysis, researchers commonly rely on tools such as DAVID~\cite{Sherman2022DAVID}, clusterProfiler~\cite{Wu2021clusterProfiler}, or Enrichr~\cite{Chen2013Enrichr}, which apply statistical methods (e.g., hypergeometric testing) to identify enriched terms. These tools generate output tables listing significant GO terms alongside adjusted p-values, fold enrichment, and associated genes. While biologically informative, such results often pose two major challenges: (1) the output can be overwhelmingly long, sometimes consisting of hundreds of terms for large gene sets, and (2) the terms are frequently redundant, with overlapping entries (e.g., ``immune response'' and ``regulation of immune response'') that fragment biological interpretation. Tools like REVIGO~\cite{Supek2011REVIGO} attempt to address these issues by clustering semantically similar terms, but they rely on fixed algorithms that cannot adapt to user-defined biological themes or align with the specific focus of a study. To overcome this limitation, some researchers manually annotate GO terms into custom functional macro-categories. While this enables hypothesis-driven interpretation, the process is labor-intensive, subjective, and poorly reproducible across annotators.

In recent years, large language models (LLMs)~\cite{achiam2023gpt,yang2025qwen3,guo2025deepseek} have emerged as powerful tools capable of understanding and generating human-like language. Trained on massive corpora, LLMs can interpret natural language input, draw on contextual and semantic relationships, and generate structured or free-form responses to a wide range of queries. These capabilities open new opportunities for tasks that involve interpreting and organizing complex biological information---such as GO term annotation~\cite{AD_biodomain}---where flexibility, contextual understanding, and scalability are critical.

In this study, we developed a \emph{training-free} and \emph{customizable} AI-enhanced framework that leverages the capabilities of LLMs to systematically annotate Gene Ontology (GO) terms into user-defined biological domains (BioDomains). We refer to this framework as \textbf{\mname} (\textbf{L}arge \textbf{L}anguage \textbf{M}odel for \textbf{B}iological \textbf{D}omains Oriented \textbf{C}lustering of Gene Ontology). 
The key insight behind \mname is to cast the GO-to-BioDomain annotation task as a textual semantic ranking problem~\cite{lu2023HiPrompt,xie2024PromptLink}. \mname first adaptively generates a prompt describing the request to return a ranked list of the most appropriate BioDomains to each query GO term. \mname then harnesses the reasoning capabilities of LLMs to perform the ranking based on semantic relevance and prior biomedical knowledge.

To evaluate its performance, we applied the proposed \mname to two distinct expert-annotated datasets: one focused on Alzheimer's disease (AD), a neurodegenerative disorder, containing 7,120 GO terms annotated to 19 BioDomains according to Gregory et al.~\cite{AD_biodomain}; and another focused on fragile X syndrome (FXS), a genetic neurodevelopmental disorder, containing 516 GO terms, which we manually annotated to 21 BioDomains. We demonstrated that LLM-driven annotation outperforms baseline methods as well as other models, in both accuracy and consistency. These results support the conclusion that LLMs represent a reliable, efficient, and user-friendly solution for GO-to-BioDomain annotation.
Together, our findings show that LLM-based annotation provides a scalable and generalizable approach for organizing GO enrichment results into biologically meaningful domains. By reducing manual effort, enhancing reproducibility, and enabling flexible, context-specific interpretation, \mname bridges the gap between statistical outputs and biological insight. Its ease of use and adaptability make it broadly applicable across diverse research areas involving gene set analysis. More broadly, this work lays the foundation for integrating LLM-driven reasoning into functional genomics workflows, offering a powerful new paradigm for interpreting high-throughput data in a biologically coherent and researcher-driven manner.

\begin{figure*}[htbp!]
  \centering
  \includegraphics[width=\linewidth]{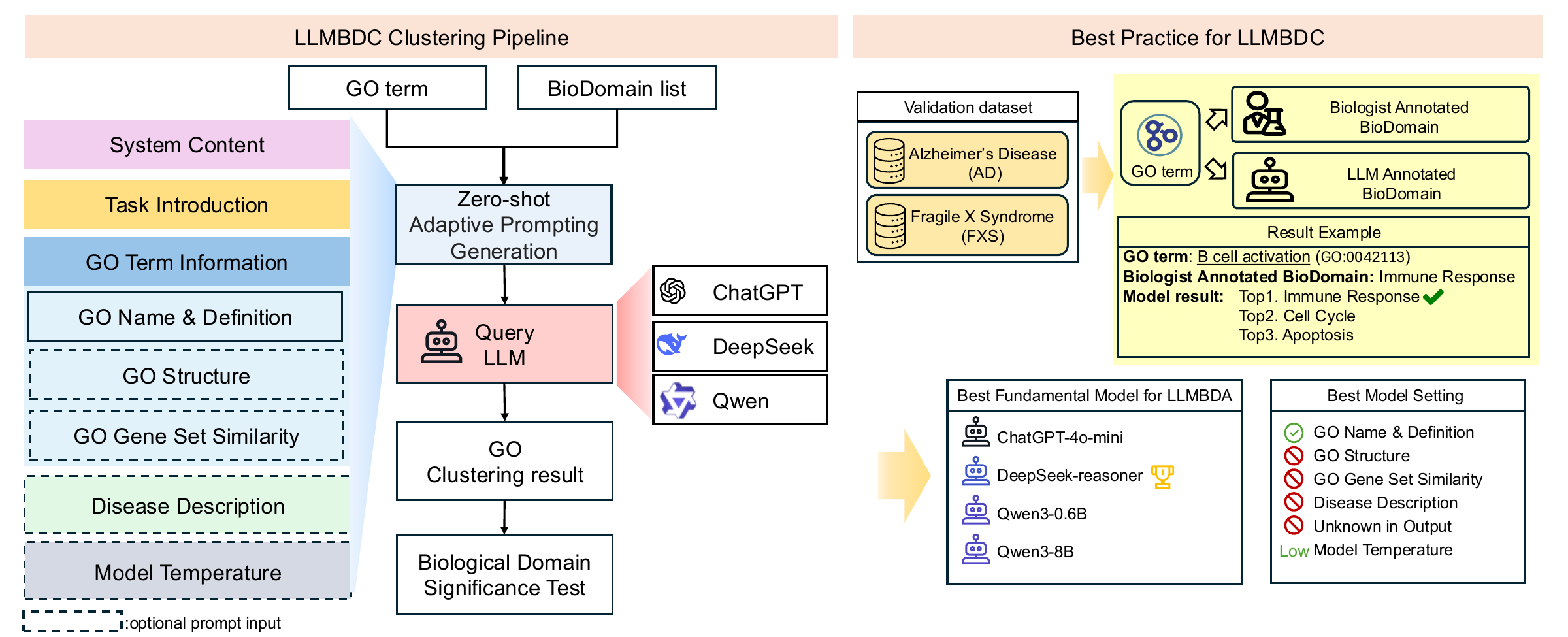}
  \vspace{-6pt}
  \caption{Overview of \mname framework.}
  \label{fig:framework_overview}
\end{figure*}

%% file: sections/related_work.tex
\section{Related Work}

\subsection{Biological Domain Annotation}

Biological domain (BioDomain) annotation offers a systems-level approach to grouping detailed GO term annotations into broader functional categories, which is especially useful for complex diseases like Alzheimer’s disease (AD). National Institute on Aging (NIA) and Alzheimer’s Association (AA) developed the \emph{Common Alzheimer’s and Related Dementias Research Ontology (CADRO)}~\cite{CADRO2024}, which organizes AD-relevant processes—such as inflammation, metabolism, and neuronal function, into standardized domains. Building on this framework, Gaiteri et al.~\cite{AD_biodomain} created a genome-wide pipeline that combines genetic association data, predicted variant effects, and multi-omic analyses to rank genes and assign them to 19 curated domains using manually selected GO terms. While effective, this method relies on static mappings that may become outdated or lack flexibility across diseases. 

Beyond these curated domain mappings, enrichment analyses often produce long lists of functionally redundant or overlapping pathways, making interpretation challenging. To address this issue, several computational methods have been developed to cluster or simplify enrichment results, thereby improving clarity and biological interpretability. These methods fall into three main categories.
First, \textbf{semantic similarity}-based methods leverage the structure of the GO graph and information‑content metrics to compute pairwise similarity between terms and remove redundancy accordingly. Representative tools include \texttt{REVIGO}~\cite{Supek2011REVIGO}, \texttt{GOSemSim}~\cite{yu2010GOSemSim}, and the \texttt{simplify()} function in \texttt{clusterProfiler}~\cite{Wu2021clusterProfiler}.
Second, \textbf{gene-set similarity} approaches shift focus from term-level semantics to the biological overlap of underlying gene sets. Methods such as \texttt{Metascape}~\cite{zhou2019metascape} compute functional or gene-overlap similarity between the gene sets associated with enriched terms, providing an alternative consolidation strategy.
Third, \textbf{ontology-structure-driven} tools reduce complexity by leveraging the hierarchical nature of the ontology itself. For example, \texttt{GO-slim}~\cite{sessa2023gogetter} maps specific GO terms to a curated subset of broader, high-level categories, thus offering a compressed but informative summary of enrichment results.
While these computational methods offer valuable complexity reduction, they also present significant limitations in practice. The number and size of clusters generated can be highly variable and often suboptimal for interpretation. For instance, \texttt{REVIGO} accepts only up to 2,000 GO terms at a time and, when given around 500 enriched pathways, may still generate over 200 clusters. Such fragmentation limits the utility of the output, as the user is left with a large number of small or overly specific clusters that require further manual interpretation. Moreover, the granularity and stability of clusters are sensitive to parameter settings and method choice, leading to inconsistent results across different analyses.

\subsection{LLMs in Gene Ontology Clustering}
Gene Ontology (GO) clustering can be viewed as a specialized instance of biomedical text summarization, in which fine-grained ontology terms must be condensed and organized into coherent higher-level functional groupings. Large language models (LLMs) have demonstrated strong potential on this broader class of biomedical ontology annotation tasks, and progress has gradually narrowed from generic biomedical representation learning toward GO-specific reasoning.

Early efforts were driven by domain-specialized pretrained LLMs that learn dense semantic biomedical representations. Models such as SapBERT~\cite{liu2020sapbert}, BioGPT~\cite{luo2022biogpt}, and PubMedBERT~\cite{gu2022pubmedbert} provide foundational embeddings that are widely reused in downstream ontology tasks, and have been further specialized for GO through structure-aware variants. GoBERT~\cite{miao2025gobertgeneontologygraph}, for instance, couples BERT with the GO graph via two GO-informed pre-training objectives, jointly capturing explicit and implicit relations among gene functions. While effective, such specialization typically demands large-scale supervised corpora~\cite{UniEntrezDB}, which are scarce in the supervision-sparse regimes characteristic of curated biological domains.

A second line of work moves away from fine-tuning toward prompting and agentic reasoning—paradigms well aligned with the summarization view of GO clustering, yet so far underexplored for biomedical ontology annotation. Prompt-based methods such as HiPrompt~\cite{lu2023HiPrompt} achieve accurate normalization of clinical terms without parameter updates. Multi-agent frameworks extend LLMs from term-level prediction to gene-set–level functional summarization: GeneAgent~\cite{wang2024geneagent} and the large-scale evaluation of \cite{hu2025eval} show that LLMs can generate coherent functional summaries directly from groups of genes. Despite these advances, none of these approaches has been adapted to the specific summarization problem of grouping GO terms into custom, user-defined biological domains—the central task we address.

We close this gap by framing GO-to-BioDomain assignment as a zero-shot biomedical annotation task built on top of GO term clustering: GO terms are first grouped into semantically coherent clusters, and an LLM is then prompted to identify the closest biological theme for each cluster and align it with a higher-level BioDomain. Both the clustering-level theme inference and the term-level domain assignment are carried out through the same LLM-driven annotation process, removing the need for static training corpora or pre-defined term–domain mappings.



%% file: sections/methods.tex
\section{Methods}
\subsection{Problem setup}
\label{ssec:problem_setup}

Enrichment analyses often produce long lists of functionally redundant or overlapping pathways, making it difficult to identify overarching biological themes.  
In this study, we consider the general task of \textit{GO clustering over biological domains}, where the goal is to map enriched GO terms into a set of broader, human-interpretable biological themes. Formulating this as a domain assignment task enables consistent comparison, flexible modeling, and statistical summarization at the domain level.

\begin{definition}[GO Clustering over biological domains]
Given a term $t_i$ from a domain-specific ontology (e.g., Gene Ontology) and a list of user-specified biological domains $\mathbf{D} = \{d_1, d_2, \dots, d_{|\mathbf{D}|}\}$, GO clustering of biological domains aims to assign the most appropriate domain labels $(\hat{d}_1, \hat{d}_2, \dots)$ to $t_i$.
\end{definition}

\subsection{\mname: Large Language Model for Biological Domains Oriented Clustering of Gene Ontology}

Following the problem definition, \mname~$f_{\theta}(\cdot)$ treats the clustering problem as a ranking task. Specifically, given a GO term $t$ and a set of candidate biological domains $\mathbf{D}$, the model returns a ranked list of domain assignments: 
\begin{equation}
    (\hat{d}_1,\hat{d}_2,\dots,\hat{d}_k) = f_{\theta}(t,\mathbf{D}),
\end{equation}
where the model parameters $\theta$ are obtained from released model weights and require no additional fine-tuning. All code and technical details for this study are publicly available at \url{https://anonymous.4open.science/r/LLMBDC_auto-615E/}.

\subsubsection{Adaptive Prompt Generation}
\mname leverages a large language model (LLM) to perform zero-shot GO clustering over biological domains. The key insight is that LLMs can be rapidly adapted to new tasks through prompt-based inference without any training~\cite{lu2023HiPrompt,xie2024PromptLink}.
The first step in \mname is to construct a prompt $\mathfrak{P}$ from the input GO term $t$ and the candidate domains $\mathbf{D}$:
\begin{equation}
    \mathfrak{P} = \Psi(t, \mathbf{D}),
\end{equation}
where $\Psi$ denotes the prompt generation function tailored for domain-level GO clustering. We implement $\Psi$ using a template-based adaptive scheme.
Fig.~\ref{fig:zero_shot_prompt} presents an example prompt for clustering the GO term ``Meiotic Chromosome Condensation'' into biological domains relevant to Fragile X Syndrome. The hyperparameter \texttt{Top\_k} is set to 5 and can be adjusted to balance between output diversity and focus.
Beyond the basic prompt structure, we also explore enriched variants that incorporate additional contextual knowledge—such as domain-specific disease descriptions, GO term definitions, and associated gene sets—as shown in Fig.~\ref{fig:framework_overview}. We evaluate these variants empirically in Sec.~\ref{sec:exp}.

\begin{figure}[htbp!]
  \centering
  \begin{tcolorbox}[colback=white,colframe=blue]
  \textcolor{lightgray}{\# System prompt to activate relevant domain knowledge in the language model}\newline
  You are a biomedical ontology expert with deep knowledge of Gene Ontology, pathway enrichment, and biological domain classification. \newline
  \textcolor{lightgray}{\# Task Introduction}\newline
  Your task is to assign the most appropriate high-level biological domain(s) to a given GO term, based on its name and functional context.From the list of Biodomains, choose the \textbf{top 5} labels that best fit this term---ranked most-to-least appropriate. \newline
 \textbf{BioDomain Candidates}: DNA Repair; Autophagy; Cell Cycle; RNA Metabolism; Epigenetic Regulation ...\newline
\textcolor{lightgray}{\# GO Term Information }\newline
  \textbf{GO Term}: Meiotic Chromosome Condensation\newline
  \textbf{GO Definition}: Compaction of chromatin structure prior to meiosis in eukaryotic cells.\newline
\textcolor{lightgray}{\# Optional prompt input [e.g.,Disease Description] }\newline
You are a biomedical ontology expert specializing in Fragile X Syndrome (FXS) research, the most common inherited form of intellectual disability and a leading genetic cause of autism spectrum disorders. FXS is caused by CGG repeat expansion in the FMR1 gene, leading to loss of FMRP and resulting in cognitive impairment, behavioral challenges, and synaptic dysfunction. Your task is to classify Gene Ontology (GO) terms in the context of FXS pathology.
\textcolor{lightgray}{\# LLM result}\newline
  \textbf{Answer}:
  \textcolor{blue}{LLM$:$} [\colorbox{cyan}{"Cell Cycle"},"DNA Repair","Structural Stabilization","RNA Metabolism","Autophagy"]
  \end{tcolorbox}
\caption{An example of adaptive prompt with task description and test query are provided to LLM.}
\label{fig:zero_shot_prompt}
\end{figure}

\subsubsection{Zero-Shot GO Clustering over Biological Domains}
Given the generated prompt $\mathfrak{P}$, we query an instruction-tuned LLM to produce a ranked list of biological domains for a given GO term. Formally, the LLM outputs:
\begin{equation}
    W = LLM_\vartheta(\mathfrak{P}) = LLM_\vartheta(\Psi(t, \mathbf{D})),
\end{equation}
where $W$ is a sequence of words, $\Psi$ is the prompt generator, and $\vartheta$ represents the frozen parameters of the pre-trained LLM. No training or fine-tuning is required, making this process fully zero-shot.

\subsubsection{LLM Output Parsing}
The output $W = (w_1, w_2, \dots, w_l)$ is a raw textual sequence, which needs to be converted into a structured list of ranked biological domains $(\hat{d}_1, \hat{d}_2, \dots, \hat{d}_k)$. Domain names may vary in length and structure (e.g., ``Apoptosis'', ``Metal Binding and Homeostasis''). To extract a clean ranked list, we apply a parsing function $\Phi(\cdot)$:
\begin{equation}
    (\hat{d}_1, \hat{d}_2, \dots, \hat{d}_k) = \Phi(W).
\end{equation}
We improve parsing reliability by including explicit formatting instructions in the prompt (e.g., JSON-style output). If the response is not well-structured, $\Phi$ uses fallback rules such as keyword matching and text normalization.

\subsubsection{Statistical Aggregation of Enrichment Scores by Domain}
After clustering GO terms into biological domains using the LLM, we aggregate their statistical significance at the domain level. Instead of reporting p-values per term, we summarize them per domain cluster.
We use the \textit{Cauchy Combination Test} (CCT)~\cite{liu2020cauchy}, which is suitable for combining dependent p-values. Given a cluster $C_j$ with $p$-values $\{p_1, p_2, \dots, p_n\}$, the test computes:
\begin{equation}
   T_j = \sum_{i=1}^{n} \tan\left\{ \left(0.5 - p_i \right) \pi \right\}, 
\end{equation}

\begin{equation}
    P_j = \frac{1}{2} - \frac{1}{\pi} \arctan(T_j).
\end{equation}
This produces a single, robust enrichment p value  $P_j$  that summary all the pathways in the cluster. We use the p value here for each domain, avoiding assumptions of independence across terms. When compared with expert-curated domain annotations, our aggregated scores align well, supporting the biological relevance of the clustering.

%% file: sections/results.tex
\section{Experimental Results}
\label{sec:exp}
\subsection{Benchmark Dataset Description}
\begin{table}[htbp!]
\centering
\caption{Benchmark Datasets overview.}
\label{tab:dataset}
\begin{tabular}{ccc}
\toprule
\textbf{Category\textbackslash Dataset} & \textbf{AD} & \textbf{FXS} \\
\midrule
\#GO Terms & 7,120 & 516\\
\#BioDomains & 19  &  21\\
\#Avg Domain/GO & 1.04 & 1 \\
\#GO w/ 1 Domain & 6,834 & 516 \\
\#GO w/ 2 Domain & 286 & 0 \\
\bottomrule
\end{tabular}
\end{table}

\begin{table*}[htbp!]
  \caption{Performance on AD and FXS datasets, reported as mean $\pm$ standard deviation (in \%).}
  \label{tab:combined_perf_dataset_column}
  \centering
  \begin{tabular}{l|l|cccc}
    \toprule
    \textbf{Dataset} & \textbf{Model} & \textbf{Precision@1} & \textbf{Precision@2} & \textbf{Recall@1} & \textbf{Recall@2} \\
    \midrule

    \multirow{7}{*}{AD}
      & Random               & $5.54 \pm 0.40$   & $5.48 \pm 0.13$   & $5.36 \pm 0.43$   & $10.57 \pm 0.28$ \\
      & GO Similarity        & $5.72 \pm 0.23$   & $5.48 \pm 0.15$   & $5.51 \pm 0.24$   & $10.55 \pm 0.34$ \\
      & GO Similarity Search & $20.48 \pm 0.12$  & $12.93 \pm 0.19$  & $20.03 \pm 0.12$  & $25.17 \pm 0.34$ \\
     & Graph Traversal      & $20.93 \pm 0.56$  & $12.84 \pm 0.30$  & $20.25 \pm 0.52$  & $24.82 \pm 0.54$ \\
     & EditDist       & $8.55 \pm 0.00$   & $7.25 \pm 0.00$   & $8.26 \pm 0.00$   & $13.97 \pm 0.00$ \\
     & SapBERT        & $48.67 \pm 0.00$  & $31.90 \pm 0.00$  & $47.16 \pm 0.00$  & $61.52 \pm 0.00$ \\
    & \textbf{\mname}      & $\mathbf{83.87 \pm 0.27}$ & $\mathbf{46.80 \pm 0.10}$ & $\mathbf{83.87 \pm 0.27}$ & $\mathbf{93.60 \pm 0.19}$ \\
    \midrule

    \multirow{7}{*}{FXS}
       & Random               & $1.04 \pm 0.23$   & $1.04 \pm 0.23$   & $1.04 \pm 0.23$   & $2.08 \pm 0.45$ \\
      & GO Similarity        & $1.43 \pm 0.68$   & $1.27 \pm 0.42$   & $1.43 \pm 0.68$   & $2.53 \pm 0.85$ \\
      & GO Similarity Search & $8.25 \pm 0.23$   & $6.50 \pm 0.41$   & $8.25 \pm 0.23$   & $13.00 \pm 0.81$ \\
      & Graph Traversal      & $13.71 \pm 0.11$  & $7.31 \pm 0.17$   & $13.71 \pm 0.11$  & $14.62 \pm 0.34$ \\
      & EditDist       & $3.12 \pm 0.00$   & $3.22 \pm 0.00$   & $3.12 \pm 0.00$   & $6.43 \pm 0.00$ \\
       & SapBERT        & $38.21 \pm 0.00$  & $25.93 \pm 0.00$  & $38.21 \pm 0.00$  & $51.85 \pm 0.00$ \\
      & \textbf{\mname}      & $\mathbf{78.41 \pm 0.25}$ & $\mathbf{46.54 \pm 0.26}$ & $\mathbf{78.41 \pm 0.25}$ & $\mathbf{93.08 \pm 0.52}$ \\
    \bottomrule
  \end{tabular}
\end{table*}

To evaluate the performance of \mname, we employed two datasets curated by expert biologists. These datasets serve as ground truth for BioDomain annotation. Table~\ref{tab:dataset} provides an overview of these two datasets. Specifically,
\begin{itemize}[leftmargin=*]
    \item \textbf{AD:} The first dataset consists of GO term annotations related to Alzheimer's disease (AD) risk, sourced from prior studies~\cite{AD_biodomain}. The annotations are based on the CADRO framework~\cite{CADRO2024}, specifically covering both general aligned and drug trial classification BioDomains. A total of 19 biological domains were annotated, each associated with AD pathology. 
    \item \textbf{FXS:} The second dataset is derived from a gene set variation analysis~\cite{hanzelmann2013gsva} of Fragile X syndrome (FXS) disease. FXS is a genetic disorder characterized by intellectual disability and developmental delay. Biologists conducted functional annotation on 516 GO terms that showed significant differential between healthy controls and FXS patients. These GO terms were classified into 21 biological domains, providing a comprehensive view of the molecular disruptions associated with FXS.
\end{itemize}
Both conditions involve complex neurological dysfunction but represent distinct etiologies: AD is a late‑onset neurodegenerative disorder, while FXS is a neurodevelopmental disorder. This contrast allows us to evaluate \mname across heterogeneous biological contexts, thus assessing its generalizability beyond a single disease framework.

\subsection{Exp 1: Comparing with Existing Biological Domain Clustering Methods}
To validate the effectiveness of \mname, we compare it against a range of zero-shot biological domain annotation baselines across two benchmark datasets: Alzheimer’s Disease (AD) and Fragile X Syndrome (FXS). These baselines are grouped into three main methodological categories:

\noindent \textbf{(a) Gene-set similarity-based methods:} 
\begin{itemize}[leftmargin=*]
    \item \textbf{GO Similarity}: this method ranks biological domains by Jaccard similarity between the gene set of a GO term and those of candidate domains. If no gene overlap is found, a random domain is assigned.
    \item \textbf{GO Similarity Search}: this method extends GO similarity by incorporating a neighborhood-based strategy. For each BioDomain, a neighborhood is constructed by selecting its top 10 most similar GO terms based on gene set overlap. Domain assignment is then performed by measuring the similarity between the input GO term and these BioDomain-specific neighbors.
\end{itemize}

\noindent \textbf{(b) GO structure-based methods:} 
\begin{itemize}[leftmargin=*]
    \item \textbf{Graph Traversal}: this method maps GO terms to biological domains by tracing their ancestors in the GO hierarchy. If an ancestor matches a known domain, it is assigned; otherwise, a random domain is used. Multiple domain matches are all included. It is worth noting that we found that 9 out of 19 biological domains in the AD dataset and 7 out of 21 in the FXS dataset can be located in GO hierarchy.
\end{itemize}

\noindent \textbf{(c) GO term semantic similarity-based methods:} 
\begin{itemize}[leftmargin=*]
    \item \textbf{EditDist}~\cite{ristad2002edistdist}: this method uses Levenshtein edit distance to assign biological domains to GO terms based purely on string similarity between their names.
    \item \textbf{SapBERT}~\cite{liu2020sapbert}: this method represents a semantic embedding-based biological domain idea, where GO terms and candidate domain names are embedded using SapBERT, and cosine similarity is used to rank the candidates. Specifically, SapBERT is a pretrained Transformer model optimized for biomedical concept linking, pre-trained on PubMed-derived biomedical texts.
\end{itemize}
We also include a \textit{Random} method that assigns domains uniformly at random as a naive baseline.

Table~\ref{tab:combined_perf_dataset_column} presents the benchmarking results. Following the problem setup in Sec.~\ref{ssec:problem_setup}, we adopt the ranking based metrics including Precision@K and Recall@K as the main evaluation metrics (the higher the better).
Across both datasets, \mname with DeepSeek-reasoner clearly outperforms all compared methods in both precision and recall. Notably, classical methods relying on gene set or semantic similarity fall short in domain generalization, while \mname effectively captures biological relevance with no task-specific training.

\begin{table}[htbp!]
\centering
\caption{Domain-level comparison between expert and \mname on the AD dataset.}
\label{tab:ad_alignment}
\begin{tabular}{p{0.3\linewidth}ccp{0.1\linewidth}p{0.1\linewidth}}
\toprule
\textbf{BioDomain} & \textbf{Truth p} & \textbf{\mname p} & \textbf{Truth Rank} & \textbf{\mname Rank} \\
\midrule
Mitochondrial Metabolism        & 3.68E-10 & 5.69E-10 & \textbf{1}  & \textbf{1}  \\
Synapse                          & 8.01E-05 & 8.62E-05 & \textbf{2}  & \textbf{2}  \\
Cell Cycle                       & 3.19E-04 & 1.45E-01 & 3  & 14 \\
Endolysosome                     & 3.88E-04 & 3.84E-04 & \textbf{4}  & \textbf{4}  \\
Structural Stabilization         & 1.42E-03 & 2.11E-04 & 5  & 3  \\
Lipid Metabolism                 & 1.47E-03 & 1.35E-03 & 6  & 5  \\
DNA Repair                       & 4.17E-03 & 4.36E-03 & 7  & 8  \\
Metal Binding \& Homeostasis     & 8.26E-03 & 3.98E-03 & 8  & 7  \\
Oxidative Stress                 & 9.28E-03 & 1.19E-02 & \textbf{9}  & \textbf{9}  \\
Proteostasis                     & 9.74E-03 & 1.62E-02 & \textbf{10} & \textbf{10} \\
Myelination                      & 2.61E-02 & 2.42E-02 & \textbf{11} & \textbf{11} \\
APP Metabolism                   & 5.94E-02 & 5.50E-02 & \textbf{12} & \textbf{12} \\
Vasculature                      & 7.89E-02 & 8.21E-02 & \textbf{13} & \textbf{13} \\
Apoptosis                        & 1.90E-01 & 2.84E-01 & 14 & 16 \\
Tau Homeostasis                  & 2.11E-01 & 2.11E-01 & \textbf{15} & \textbf{15} \\
RNA Spliceosome                  & 7.81E-01 & 4.84E-01 & 16 & 17 \\
Immune Response                  & 1.00     & 1.00     & 17 & 18 \\
Epigenetic                       & 1.00     & 1.00     & 18 & 19 \\
Autophagy                        & 1.00     & 1.48E-03 & 19 & 6  \\
\bottomrule
\end{tabular}
\end{table}

\begin{table}[htbp!]
\centering
\caption{Domain-level comparison between expert and \mname on the FXS dataset.}
\label{tab:fxs_alignment}
\begin{tabular}{p{0.3\linewidth}ccp{0.1\linewidth}p{0.1\linewidth}}
\toprule
\textbf{BioDomain} & \textbf{Truth p} & \textbf{\mname p} & \textbf{Truth Rank} & \textbf{\mname Rank} \\
\midrule
Immune System and Inflammation                    & 1.09E-05 & 1.06E-05 & \textbf{1}  & \textbf{1}  \\
Epigenetic Regulation                              & 5.34E-05 & 4.75E-05 & 2  & 3  \\
Structural Stabilization                           & 6.16E-05 & 2.64E-05 & \textbf{3}  & \textbf{2}  \\
Neurotransmission and Synaptic Regulation          & 7.47E-05 & 7.47E-05 & \textbf{4}  & \textbf{4}  \\
Autophagy                                          & 1.09E-04 & 1.27E-04 & \textbf{5}  & \textbf{5}  \\
Signal Transduction                                & 2.12E-04 & 2.69E-04 & \textbf{6}  & \textbf{7}  \\
DNA Repair                                         & 3.47E-04 & 7.75E-04 & \textbf{7}  & \textbf{9}  \\
Protein Metabolism and Trafficking                 & 4.47E-04 & 5.46E-03 & 8  & 16 \\
System and Developmental Process                   & 7.51E-04 & 4.65E-04 & 9  & 8  \\
Apoptosis                                          & 1.84E-03 & 1.92E-03 & \textbf{10} & \textbf{12} \\
Response to Stimulus                               & 1.84E-03 & 1.61E-03 & 11 & 11 \\
Gliogenesis and Glial Differentiation              & 1.98E-03 & 1.97E-03 & \textbf{12} & \textbf{13} \\
Mitochondrial Function and Metabolism              & 2.34E-03 & 2.00E-03 & \textbf{13} & \textbf{14} \\
RNA Metabolism                                     & 3.58E-03 & 3.48E-03 & \textbf{14} & \textbf{15} \\
Neurodevelopment and Neuronal Differentiation      & 4.92E-03 & 5.64E-03 & \textbf{15} & \textbf{18} \\
Intracellular Trafficking and Organelle Dynamics   & 5.19E-03 & 2.45E-04 & 16 & 6  \\
Lipid Metabolism                                   & 5.31E-03 & 5.55E-03 & 17 & 17 \\
Transcription and Translation Machinery            & 5.66E-03 & 8.36E-03 & \textbf{18} & \textbf{19} \\
Molecular Transport and Homeostasis                & 6.06E-03 & 9.06E-04 & 19 & 10 \\
Cell Cycle                                         & 1.11E-02 & 2.34E-02 & \textbf{20} & \textbf{21} \\
Cell Adhesion and Interaction                      & 3.72E-02 & 2.31E-02 & 21 & 20 \\
\bottomrule
\end{tabular}
\end{table}

\subsubsection{In-depth analysis}
\begin{figure}[htbp!]
    \centering
    \includegraphics[width=0.95\linewidth]{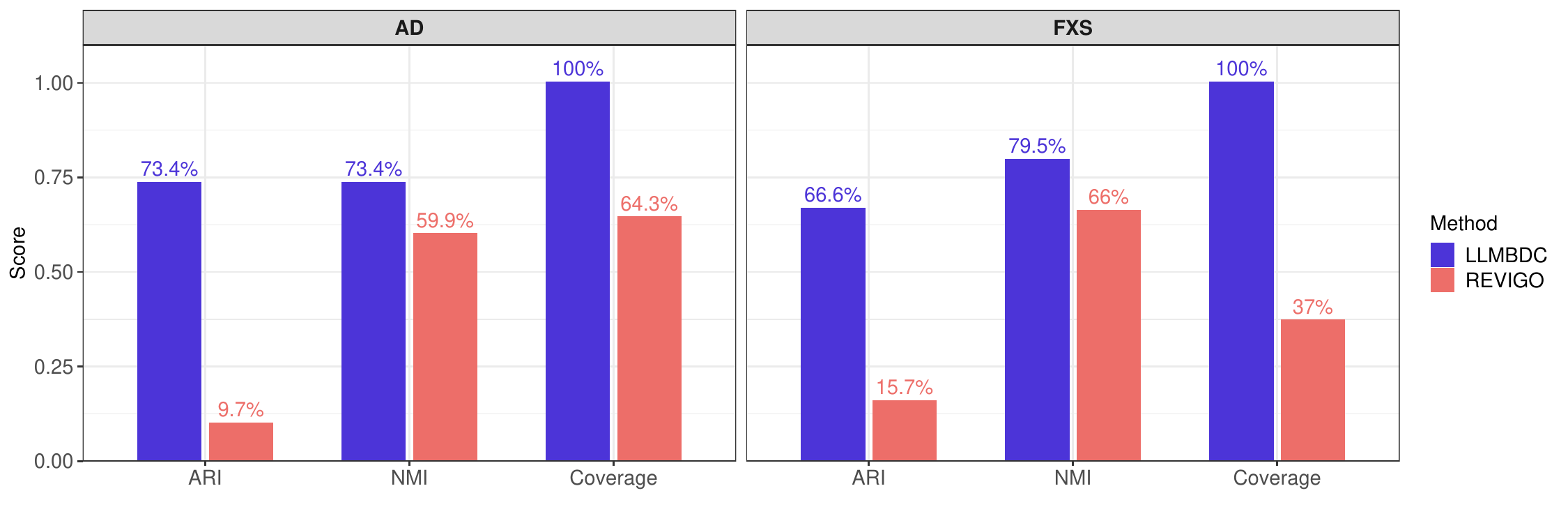}
    \vspace{-12pt}
    \caption{Domain-level signficance rank comparison between \mname and expert-based GO clustering. Most top-ranked domains by \mname are consistent with expert prioritization.}
    \label{fig:domain_p_comparison}
\end{figure}

We further assess the alignment between \mname-generated clustering and expert-defined domain-level GO enrichment using statistical evidence. Specifically, we compare the domain-level Cauchy Combination Test (CCT)~\cite{liu2020cauchy} $p$-values from \mname to those derived from expert-labeled GO clusters across two benchmark datasets: Alzheimer’s Disease (AD) and Fragile X Syndrome (FXS). 
As shown in Fig.~\ref{fig:domain_p_comparison}, the top-ranked biological domains prioritized by \mname closely match the expert-derived clusters in both datasets. To further support this finding, we present domain-level $p$-values and rankings side-by-side in Tables~\ref{tab:ad_alignment} and~\ref{tab:fxs_alignment}.
Across both datasets, \mname reliably ranks the most biologically relevant domains at the top. In AD, \textit{Mitochondrial Metabolism} and \textit{Synapse} are the top two in both expert and \mname results. In FXS, domains like \textit{Immune System}, \textit{Structural Stabilization}, and \textit{Neurotransmission} show strong agreement. While minor mismatches occur, such as in \textit{Cell Cycle} or \textit{Protein Metabolism}, the overall ranking correlation indicates that \mname effectively recovers expert-level prioritization without manual input.

\begin{table*}[htbp!]
  \caption{Performance on AD and FXS datasets, reported as mean $\pm$ standard deviation (in \%).}
  \label{tab:merged_perf}
  \centering
  \begin{tabular}{l|l|cccc}
    \toprule
    \textbf{Dataset} & \textbf{Model} & \textbf{Precision@1} & \textbf{Precision@2} & \textbf{Recall@1} & \textbf{Recall@2} \\
    \midrule
    AD  & \mname(Qwen3-0.6B)          & $19.55 \pm 0.17$ & $14.97 \pm 0.07$ & $18.72 \pm 0.17$ & $28.44 \pm 0.10$ \\
        & \mname(Qwen3-8B)            & $77.02 \pm 0.15$ & $44.20 \pm 0.05$ & $75.29 \pm 0.15$ & $85.35 \pm 0.12$ \\
        & \mname(ChatGPT-4o-mini)     & $77.39 \pm 0.68$ & $43.55 \pm 0.34$ & $77.39 \pm 0.68$ & $87.09 \pm 0.69$ \\
        & \textbf{\mname(DeepSeek-reasoner)} & $\mathbf{83.87 \pm 0.27}$ & $\mathbf{46.80 \pm 0.10}$ & $\mathbf{83.87 \pm 0.27}$ & $\mathbf{93.60 \pm 0.19}$ \\
    \midrule
    FXS & \mname(Qwen3-0.6B)          & $5.31 \pm 4.23$  & $8.16 \pm 7.93$  & $5.31 \pm 4.23$  & $16.31 \pm 1.59$ \\
        & \mname(Qwen3-8B)            & $67.77 \pm 0.23$ & $39.31 \pm 0.06$ & $67.77 \pm 0.23$ & $78.62 \pm 0.11$ \\
        & \mname(ChatGPT-4o-mini)     & $36.26 \pm 0.94$ & $23.76 \pm 0.61$ & $36.26 \pm 0.94$ & $47.51 \pm 1.22$ \\
        & \textbf{\mname(DeepSeek-reasoner)} & $\mathbf{78.41 \pm 0.25}$ & $\mathbf{46.54 \pm 0.26}$ & $\mathbf{78.41 \pm 0.25}$ & $\mathbf{93.08 \pm 0.52}$ \\
    \bottomrule
  \end{tabular}
\end{table*}

\subsection{Exp 2: Comparing with Existing GO Clustering Method}
To assess the advantages of our proposed \mname over traditional clustering approaches, we compared it against REVIGO~\cite{Supek2011REVIGO}, a widely used method for reducing redundancy in GO enrichment results. REVIGO groups semantically similar GO terms based on information‑content measures and does not rely on any external domain knowledge. In contrast, \mname leverages an LLM to assign GO terms to user‑defined biological domains, even when the domain list is not previously aligned with the ontology. To provide a fair and challenging comparison, we ran both methods under the same “unknown biodomain list” setting (\textit{i.e.,} meaning models was not given any predefined mapping), and evaluated how well their outputs match ground‑truth domain annotations.
We conducted a second benchmarking analysis based on three quantitative metrics: Adjusted Rand Index (ARI), Normalized Mutual Information (NMI), and Coverage. Using the unknown biodomain list as input, we compared the semantic clusters produced by REVIGO against the biodomain assignments generated by \mname.
Across both datasets, \mname consistently outperformed REVIGO. For the AD dataset, \mname achieved an ARI of 0.73, NMI of 0.73, and full Coverage (100\%), whereas REVIGO obtained substantially lower scores (ARI = 0.10, NMI = 0.60) and covered only 64\% of the enriched GO terms. A similar pattern was observed in FXS: \mname reached ARI = 0.67, NMI = 0.80, and 100\% Coverage, compared to REVIGO’s ARI = 0.16, NMI = 0.66, and 37\% Coverage. These results demonstrate that even without preset biodomain information, \mname produces clusters that are more consistent, information-rich, and comprehensive than those generated by 
REVIGO.


\subsection{Ablation Study}
\begin{figure*}[htbp!]
  \centering
  \begin{subfigure}{\textwidth}
    \includegraphics[width=0.95\linewidth]{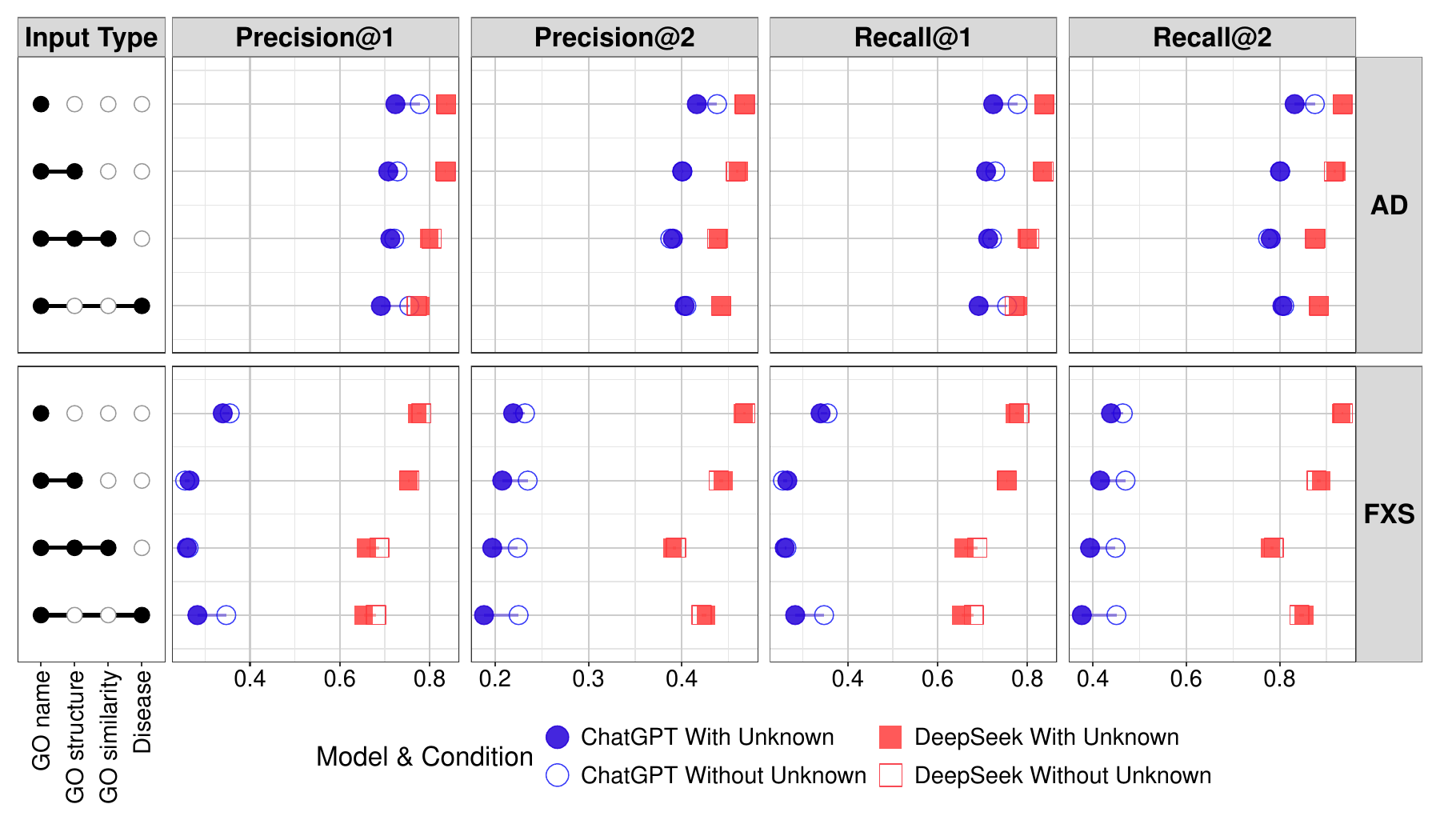}
    \vspace{-16pt}
    \caption{Unknown label effect }
    \label{fig:unknown_comp}
  \end{subfigure}
  \vspace{5pt}
  
  \begin{subfigure}{\textwidth}
    \includegraphics[width=0.95\linewidth]{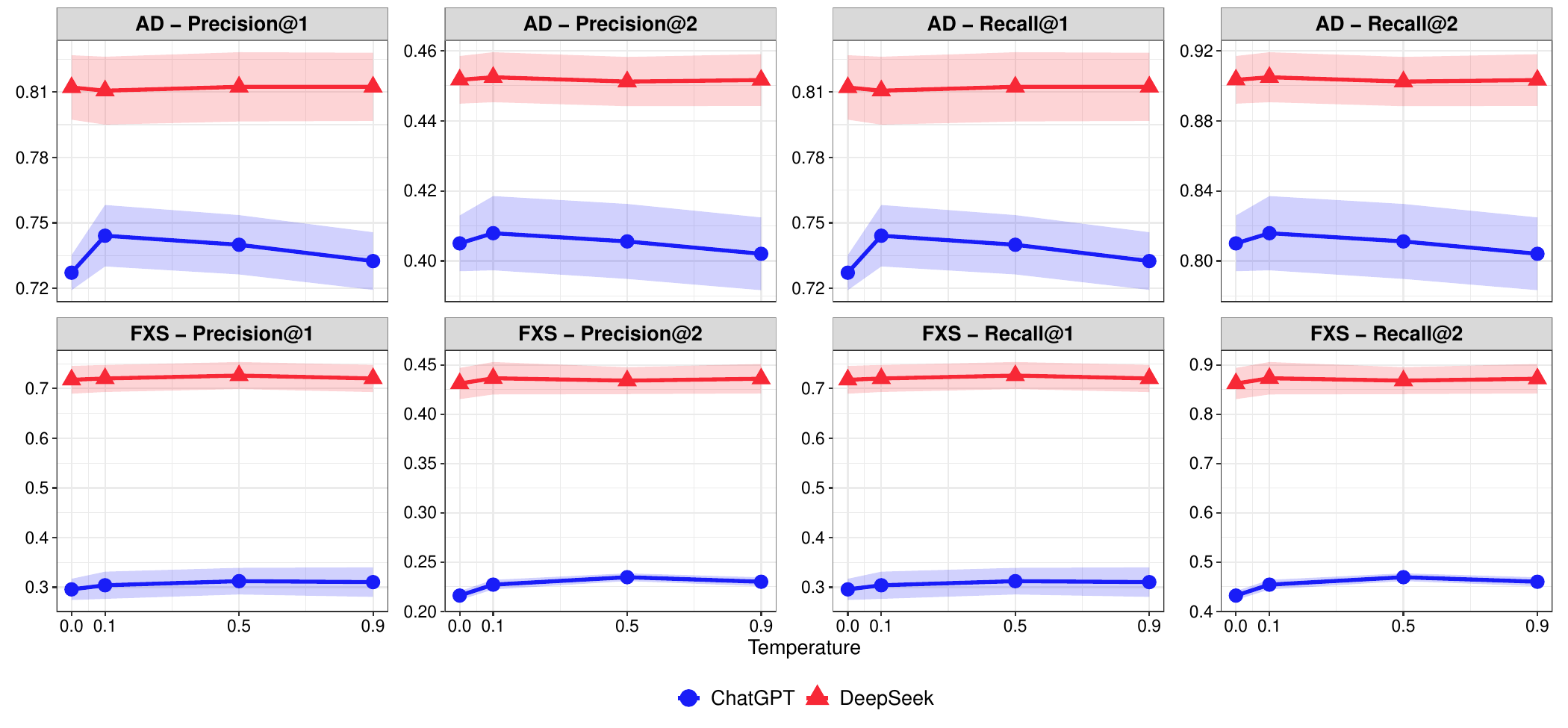}
    \vspace{-8pt}
    \caption{Decoding temperature effect}
    \label{fig:tem_comp}
  \end{subfigure}
  \caption{Effect of prompting input, decoding temperature and ``Unknown'' label option on \mname performance across AD and FXS datasets.}
  \label{fig:combined_ablation}
\end{figure*}

We systematically evaluated the optimal configuration for deploying \mname in GO clustering over biological domains using two benchmark datasets: Alzheimer’s Disease (AD) and Fragile X Syndrome (FXS), with expert annotations as reference. Our analysis focused on three factors: decoding temperature, the use of an ``Unknown'' label, and prompt content.

\noindent \textbf{Temperature and ``Unknown'' Label.}  
As shown in Fig.~\ref{fig:unknown_comp} and Fig.~\ref{fig:tem_comp}, a low decoding temperature (0.0 or 0.1) yields better precision and recall, suggesting that deterministic decoding ensures more consistent clustering. Additionally, disabling the ``Unknown'' label option improves recall by encouraging full domain assignments, even in uncertain cases.

\noindent \textbf{Prompt Input.}  
We tested four prompt input formats and found that using only the GO term name and its definition led to the best results. Adding GO structure, gene set metadata, or disease descriptions did not improve performance and sometimes reduced accuracy, likely due to added noise.

\noindent \textbf{Model Choice of LLM.}  
Table~\ref{tab:merged_perf} reports the performance of various LLM models used in \mname. We evaluate both open-source models (Qwen3-0.6B and Qwen3-8B~\cite{yang2025qwen3}) and commercial models (ChatGPT-4o-mini~\cite{achiam2023gpt} and DeepSeek-reasoner~\cite{guo2025deepseek}). DeepSeek-reasoner consistently outperforms all others, achieving the highest precision and recall on both datasets.

\noindent \textbf{Recommended Configuration.}  
Our findings suggest that the best practice for \mname is to use DeepSeek-reasoner with a low temperature (0.0–0.1), \textbf{disable the ``Unknown'' label}, and only include the GO name and definition in the prompt. This configuration provides the most accurate and interpretable clustering performance.

%% file: sections/discussion.tex
\section{Discussion}



\noindent \textbf{Flexibility and Customizability.} A key strength of \mname is its flexibility and customizability. Unlike static clustering tools such as REVIGO, which group GO terms based on fixed algorithms or ontology tree distances, \mname can incorporate custom-defined BioDomains and include optional comprehensive information---such as disease-specific descriptions or GO term hierarchy structures---into its annotation process. Currently, BioDomains are defined by domain experts, such as biologists or clinical researchers, based on prior knowledge and the specific context of the study. In the future, these BioDomains could be generated with the assistance of large language models (LLMs), reducing manual effort and improving scalability. This makes \mname well suited for researchers who wish to organize GO terms according to domain-specific biological themes, such as ``Neural Circuit Assembly,'' ``Stress Response Pathways,'' or ``Mitochondrial Function,'' depending on the study focus.
Another practical advantage of \mname lies in its ability to flexibly accommodate updates to the GO database. Because it operates through natural language reasoning rather than relying on fixed ontology structures or predefined hierarchical mappings, \mname can seamlessly adapt to changes in the GO term catalog. As the Gene Ontology continues to evolve---with new terms added, existing definitions refined, and obsolete terms removed---\mname allows researchers to incorporate the most current, or any version of the ontology, into their analyses. This flexibility reduces reliance on static annotations and ensures that downstream biological interpretations remain aligned with the latest community standards.

\noindent \textbf{Scalability, Reproducibility, and Model Selection.} 
Importantly, \mname also improves scalability and reproducibility. Manual grouping of GO terms is labor-intensive and prone to subjectivity, often leading to inconsistency between annotators. By applying consistent semantic reasoning through LLMs, \mname reduces inter-user variability and allows for reproducible annotation at scale. This feature is particularly valuable in large collaborative projects or automated bioinformatics pipelines where standardization is crucial.
We also explored how the choice of underlying LLM affects \mname's performance. Using an identical prompt structure, we compared annotations generated by different fundamental models (e.g., gpt-4o-mini, DeepSeek-Reasoning). Interestingly, DeepSeek Reasoning achieved higher annotation accuracy and closer alignment with human-labeled BioDomains. However, this improvement came at a cost: DeepSeek consumed more tokens, required longer runtime, and incurred higher computational expense than gpt-4o-mini. These differences highlight that the performance gap is not necessarily due to one model being universally superior, but may instead reflect distinctions in their reasoning depth, prompt interpretation, or architectural design. Reasoning-oriented models like DeepSeek may be particularly well-suited for tasks involving nuanced semantic ranking, but their higher cost and slower throughput may limit scalability in certain use cases. Therefore, users should weigh the benefits in annotation quality against practical trade-offs such as speed, cost, and computational resources when selecting an LLM backend for \mname.

\noindent \textbf{Limitations and Future Directions.}
While \mname shows strong performance and versatility, there are limitations to consider. LLM outputs can still be influenced by prompt phrasing, model-specific biases, or limited knowledge of rare or emerging GO terms. Moreover, although \mname allows optional contextual input, its current implementation does not yet fully leverage structured biological knowledge graphs~\cite{cui2023survey} or curated databases such as KEGG~\cite{kanehisa2023kegg}, Reactome~\cite{milacic2024reactome}, or STRING~\cite{szklarczyk2023string}. Future extensions of this framework could explore hybrid strategies that integrate LLM reasoning with structured ontologies, graph-based embeddings, or pathway co-membership information. Additionally, as larger and more specialized LLMs become available, benchmarking across domains (\textit{e.g.}, immunology, developmental biology, or cancer) will be important for optimizing annotation performance.


\noindent \textbf{Ethical Considerations.} \mname relies on large language models that may inherit biases from their training data, potentially leading to annotation inaccuracies for underrepresented or emerging biological processes. Users should exercise caution when interpreting outputs for poorly characterized GO terms or novel domains where the LLM lacks sufficient knowledge. All data used in this study (\textit{e.g.}, GO term annotations and curated BioDomain labels) are derived from published, fully de‑identified sources and comply with HIPAA standards, as no human subject or personally identifiable information was involved. Additionally, while \mname improves reproducibility by automating annotation, it does not eliminate the need for expert review, especially in high‑stakes settings such as clinical target discovery or drug development. Transparent reporting of the underlying LLM version, prompt design, and any optional context is essential to ensure result interpretability and cross‑study comparability.

%% file: sections/conclusion.tex
\section{Conclusion}


We introduced \textbf{\mname}, a flexible, zero-shot framework that leverages large language models to translate Gene Ontology (GO) terms into biologically meaningful, user-defined BioDomains. By framing GO-to-BioDomain annotation as a semantic ranking problem, \mname avoids costly training and manual curation while enabling context-aware, interpretable grouping of GO terms. We evaluated \mname on two biologically distinct, manually curated datasets, which differed in GO term counts and BioDomain definitions. Across both settings, \mname consistently outperformed traditional GO-based methods and alternative models like SapBERT, achieving higher accuracy and semantic alignment with human annotations. These results demonstrate strong generalizability across diverse biological systems and research contexts. By reducing redundancy, improving reproducibility, and adapting seamlessly to user-defined annotation schemes, \mname bridges the gap between statistical enrichment outputs and higher-level biological interpretation. It empowers researchers to extract clearer insights from large-scale gene sets, accelerate hypothesis generation, and stay aligned with evolving GO standards. These capabilities make the proposed \mname a practical, interpretable tool for modern functional genomics analysis.